# Evolution of Weyl orbit and quantum Hall effect in Dirac semimetal $Cd_3As_2$

Cheng Zhang[1,2], Awadhesh Narayan[3,4], Shiheng Lu[1,2], Jinglei Zhang[5], Huiqin Zhang[1,2], Zhuoliang Ni[1,2], Xiang Yuan[1,2], Yanwen Liu[1,2], Ju-Hyun Park[6], Enze Zhang[1,2], Weiyi Wang[1,2], Shanshan Liu[1,2], Long Cheng[5], Li Pi[5], Zhigao Sheng[5], Stefano Sanvito[3], Faxian Xiu[1,2,7]*

[1] State Key Laboratory of Surface Physics and Department of Physics, Fudan University, Shanghai 200433, China

[2] Collaborative Innovation Center of Advanced Microstructures, Nanjing 210093, China

[3] School of Physics, CRANN and AMBER, Trinity College, Dublin 2, Ireland

[4] Department of Physics, University of Illinois at Urbana-Champaign, Illinois, USA

[5] High Magnetic Field Laboratory, Chinese Academy of Sciences, Hefei 230031, China

[6] National High Magnetic Field Laboratory, Tallahassee, Florida 32310, USA

[7] Institute for Nanoelectronic Devices and Quantum Computing, Fudan University, Shanghai 200433, China

* Correspondence and requests for materials should be addressed to F. X. (E-mail: Faxian@fudan.edu.cn)

**Abstract**

Owing to the coupling between open Fermi arcs on opposite surfaces, topological Dirac semimetals exhibit a new type of cyclotron orbit in the surface states known as Weyl orbit. Here, by lowering the carrier density in $Cd_3As_2$ nanoplates, we observe a crossover from multiple- to single-frequency Shubnikov–de Haas (SdH) oscillations when subjected to out-of-plane magnetic field, indicating the dominant role of surface transport. With the increase of magnetic field, the SdH oscillations further develop into quantum Hall state with non-vanishing longitudinal resistance. By tracking the oscillation frequency and Hall plateau, we observe a Zeeman-related splitting and extract the Landau level index as well as sub-band number. Different from conventional two-dimensional systems, this unique quantum Hall effect may be related to the quantized version of Weyl orbits. Our results call for further investigations into the exotic quantum Hall states in the low-dimensional structure of topological semimetals.

**Introduction**

The concept of topological order has strongly influenced the traditional classification of phases in condensed matter, from insulators to metals.[1, 2, 3] Being one of the most active research fields in physics, topological materials represent one type of systems that manifests unconventional boundary states, distinct from the bulk.[1, 2, 3] Recently, another important class of topological materials, Weyl semimetals, has been discovered.[4, 5, 6] The quasiparticles in their low-energy band dispersions, as an analogue of the long-sought Weyl fermions in particle physics, have attracted a rapidly growing research interest.[3, 4] In a Weyl semimetal, the conduction and valence bands touch each other only at several discrete points (Weyl nodes) in the momentum space.[4, 7] Acting as monopoles for the Berry flux, these Weyl nodes always exist in pairs with the opposite chirality to ensure a zero net flux.[4] Different from topological insulators, the surface states of Weyl semimetals are arc-like without forming a closed loop or reaching the boundary of the Brillouin zone, and thus they are called Fermi arcs.[4, 7] The surface Fermi-arc dispersions merge into the bulk Weyl node pairs at both ends of the arcs.[4, 7]

As widely studied earlier, the surface states of topological insulators have been found to host many interesting phenomena, ranging from quantum anomalous Hall effect[8] to high-efficiency spin-orbit transfer torque[9]. In contrast, the study of topological semimetals so far has been mainly focused on the bulk properties.[10, 11, 12, 13, 14] On-going experimental investigations of Fermi arcs are limited mostly to spectroscopic methods, like angle-resolved photoelectron spectroscopy (ARPES)[5] or scanning tunneling microscope (STM)[15]. Transport measurements as a tool to study such Fermi arcs have been challenging. Weyl semimetals are gapless and the bulk states might dominate the transport even with the Fermi level lying near the Weyl nodes.[4, 7] Signature of a conducting surface channel in topological semimetals was found through the detection of Aharonov-Bohm oscillations.[16] Unlike the surface states of topological insulators, Fermi arcs are open orbits and will not lead to conventional Shubnikov–de Haas (SdH) oscillations.[7] However, as pointed out by Potter *et al.*,[17] a complete cyclotron orbit can be developed if the Fermi arcs on the opposite surfaces can couple together to form a closed loop. To accomplish such coupling, the overall thickness of the sample should be small compared to the quantum mean free path so that electrons can directly propagate between two surfaces via the bulk state without destroying the phase coherence.[17] Hence, this novel "Weyl orbit", involving electron cyclotrons in the momentum space and a tunneling process in real space, will only appear in Weyl semimetals with low dimensionality, while being absent in the bulk form.[17] Such correlation with bulk states is a distinct feature of the Weyl orbit. Experimental evidence of the Weyl orbit has only been provided for $Cd_3As_2$ microflakes prepared by focused ion beam (FIB) method.[18] It is noteworthy that $Cd_3As_2$ is a Dirac semimetal, which can be viewed as a special Weyl semimetal with opposite Weyl node pairs overlapping in the momentum space.[17, 19] An additional two-dimensional (2D) Fermi surface emerges[18] when the thickness of the sample is below ~1 μm. This demonstrates the feasibility of investigating SdH oscillations as an effective mean for studying the Weyl orbit.

In this study, we report the evolution of 2D surface transport and quantum Hall effect (QHE) of Dirac semimetals when approaching the quantum limit based on a series of $Cd_3As_2$ nanoplates. By decreasing the Fermi level, we observed a transition from the coexistence of bulk and surface SdH oscillations to surface-dominant transport under out-of-plane magnetic field. Further increasing the field drives the system into the quantum Hall state. By tracking the oscillation frequency and Hall plateau, we observe a Zeeman-related splitting behavior under high field and extract Landau level (LL) index as well as the sub-band number. Several possible scenarios for the origin of the quantum Hall effect are discussed. The emergence of QHE in $Cd_3As_2$ nanostructures opens up a new frontier in the research of topological semimetals.

**Results**

**Additional 2D Fermi surface**

The simple bulk band structure and the controllable growth of nanoplates with different Fermi levels make $Cd_3As_2$ a good candidate for studying the surface states through SdH oscillations.[13, 20, 21, 22] A standard fabrication process was applied to make electrical contact to the as-grown nanoplates (a typical thickness around 80~150 nm, Supplementary Fig. 1). Magnetotransport with rotational field direction is conducted to detect the SdH oscillations of Weyl orbit in $Cd_3As_2$ (refer to Fig. 1a and 1b for illustration). Figure 1c shows a typical magnetoresistance (MR) curve of $Cd_3As_2$ nanoplate (sample #1) with perpendicular magnetic field *B* at 2 K. Clear beating patterns are observed in the extracted SdH oscillations (the inset of Fig. 1c), indicating possible multiple cyclotron orbits. This behavior is in stark contrast with the single-frequency SdH oscillations detected in bulk $Cd_3As_2$ crystals grown by the same method (Fig. 1d). Figure 1 e and f present the corresponding fast Fourier transformation (FFT) spectra of the oscillations in sample #1 and bulk sample, respectively. Two peaks are found in the FFT spectrum of the nanoplate sample while only one distinct peak exists in that of the bulk sample. As previously reported,[10, 14] the SdH oscillations in $Cd_3As_2$ bulk crystals mostly exhibit single oscillation frequency with nearly isotropic Fermi surface. Nevertheless, the beating pattern of SdH oscillations in $Cd_3As_2$ has also been suggested to result from Fermi surface nesting[23] or band curvature near the Lifshitz transition[24] in bulk states. Notably, the bulk states in both samples show a similar oscillation frequency, indicating very close Fermi levels in these samples. This excludes the influence of band structure difference at different Fermi levels. In order to further investigate its origin, we map out the angle dependence of the SdH oscillations. The two insets in Fig. 1 e and f are the angle dependence of the oscillation frequency $F$ with magnetic field rotating from the out-of-plane ($\theta=0°$) to the in-plane ($\theta=90°$) direction (refer to Supplementary Fig. 2 for the angle dependence of SdH oscillations). Except for the isotropic frequency originating from the bulk Fermi surface, the other oscillation frequency increases as the field rotates away from the perpendicular direction, while its amplitude decreases rapidly and finally cannot be clearly resolved at large angles (Supplementary Fig. 2). The angle dependence of the frequency can be well fitted by $1/\cos\theta$ function, corresponding to the 2D Fermi surface. The consistency of the 2D Fermi surface in as-grown $Cd_3As_2$ nanoplates helps us to eliminate the concern of the trivial defect layers from the FIB fabrication process as a possible origin for the new oscillation frequency in previous experiments[18].

In order to study how the Weyl orbit evolves with Fermi level, we further select Cd$_3$As$_2$ nanoplates with low carrier density (on the order of $10^{17}$ cm$^{-3}$). Figure 2a-b displays the angle dependence of the SdH oscillations and the LL spectra of sample #2 at 2.5 K, respectively. For clarity, the MR background in Fig. 2a is removed (refer to Supplementary Fig. 3 for the original MR and Hall data). The carrier density is calculated to be $n = 5.7 \times 10^{17}$ cm$^{-3}$ from Hall coefficient, which is an order of magnitude lower than that in conventional bulk Cd$_3$As$_2$ crystals grown by flux method[14]. The LL spectra in Fig. 2b show both the peak and valley positions of the SdH oscillations. The dimensionality of the Fermi surfaces is distinguished by their angle dependence. Different from the case in high-Fermi-level samples, the three-dimensional (3D) oscillations cannot be resolved in MR at small angles. Apparently, the oscillations at $\theta$=90° are not comparable in amplitude to that of the oscillations at $\theta$=0° unless they are enlarged by a factor of 20 (Fig. 2a). By tracking the LL spectra with magnetic field (Fig. 2b), an anomalous feature becomes evident that the periodicity clearly changes at large fields. To investigate its origin, firstly we fit the frequency in the low field regime as shown in the Landau fan diagram (Fig. 2c). Based on the oscillations at $\theta$=90°, the $N$=1 bulk LL is achieved at around 0.28 T$^{-1}$, as marked by the black dash line. Similar to sample #1 (Fig. 1e inset), the low field oscillation frequency fits well with the 2D Fermi surface (Fig. 2d), which gives a strong evidence for the persistence of surface states in low-Fermi-level samples. Then we went back to check the oscillations at high fields. As plotted in Fig. 2e and Supplementary Fig. 4, both $\Delta R$ and $-\frac{d^2 R}{dB^2}$ are used to extract the LL positions. Surprisingly, two hidden peaks are revealed by the $-\frac{d^2R}{dB^2}$ plot, while they appear less obvious in the $\Delta R$ curve. Consequently, in the regime of 4~9 T, there is a splitting-like behavior emerging in the SdH oscillations that induce the increase of periodicity, as marked by the green and magenta arrows in Fig. 2e, respectively. Similar behavior happens in other samples as well (Supplementary Fig. 5 and Supplementary Note 1). It is noted that here the bulk state of sample #2 has reached the quantum limit at the corresponding field regime, where the bulk carriers are already confined to the lowest LL and can no longer produce any quantum oscillations.

**Quantum Hall effect**

To further investigate the change of the oscillation periodicity, a higher magnetic field up to 18 T was applied as shown in Fig. 3a-b. Consistent with sample #2, multiple pronounced SdH oscillations emerge with the increase of the magnetic field (Fig. 3b). Strikingly, quantized plateaus appear in the Hall resistance $R_{xy}$ of sample #7 in Fig. 4a. Supplementary Figure 6 also exhibits another set of data (sample #6), in which the Hall plateau is accompanied by a nearly flat valley in the longitudinal MR, thus providing a clear evidence for QHE at high magnetic fields. This kind of quantum Hall samples usually has a high carrier mobility (~$10^5$ cm$^2$/Vs) at low temperatures. By converting the Hall resistance to quantum resistance, we have marked the filling factors $\nu$ of the LLs corresponding to each oscillation as indexed in Fig. 3a. The change of filling factor between adjacent oscillations alters from 4 to 2 after the splitting at high fields. Such a splitting behavior could come from the Zeeman-related effect induced by external magnetic field[14, 25] (see Supplementary Note 2 for the detailed analysis). The splitting becomes more pronounced when the oscillations are plotted against $1/B$ (Fig. 3b, black arrows). We also extract the Landau fan diagram based on the oscillation peak position in the inset of Fig. 3b to determine the LL index. Here we use the center of the splitted peaks as the original LL positions.

With the increase of temperature, the oscillation minima in $R_{xx}$ gradually increase (Fig. 4a), from which we can determine the energy gap $\Delta E$. For a fixed magnetic field, the energy gap of LL with a filling factor of $2n$ is given by $\Delta E_{even} = E_c - E_z - \Gamma$ when $n$ is an even number and $\Delta E_{odd} = E_z - \Gamma$ when $n$ is an odd number.[26] Here $E_c$ is the cyclotron energy, $E_z$ is the Zeeman energy, and $\Gamma$ is the LL broadening factor caused by disorder and scattering. The minima of $R_{xx}$ follow a typical thermal activation relation described by the formula $R_{xx} \propto e^{-\Delta E/2k_B T}$, where $k_B$ is the Boltzmann constant. Performing linear fits to the Arrhenius plot of $\ln R_{xx}^{min}$ as a function of $1/T$ (Fig. 4c) yields the activation energy of 4.0 K and 3.4 K for $\nu = 10$ and 8 in sample #7, as well as 2.5 K and 10.3 K for $\nu = 20$ and 10 in sample #6. Note that the energy gap for $\Delta E_{odd}$ is quite large compared to that of $\Delta E_{even}$, suggesting a significant energy splitting from Zeeman effect. It can also be confirmed by the weak oscillation amplitude for $\nu = 16$ in both samples since $\Delta E_{even}$ becomes smaller as the Zeeman energy gets larger.

When encountering size confinement along certain dimension, 3D electronic states will evolve into discrete energy levels, *i.e.* sub-bands. If the size confinement is strong enough, it may lead to a phase transition from Dirac semimetal to quantum spin Hall insulator along with a gap forming in the bulk states.[19] To determine whether the quantum confinement is significant in our samples, we measured the longitudinal resistance $R_{xx}$ with the magnetic field applied parallel to the sample surface as shown in Fig. 5a. Clear SdH oscillations of bulk states can be observed with a frequency of 16.7 T, which indicates the successful formation of continuous cyclotron orbits within the cross-section of nanoplates. Typically, low-carrier-density systems such as narrow-gap semiconductors or semimetals only show significant quantum confinement effect below its Fermi wavelength.[27, 28] On the other hand, from the SdH oscillation frequency, we can estimate the Fermi wave vector as $k_F = 0.015$ Å through $2\pi k_F^2 = S_F = 2\pi^2 F/\phi_0$ with $\phi_0 = h/2e$. Here we regard the Fermi surface as two slightly overlapping spheres as discussed in Supplementary Note 3. Then we can obtain the Fermi wavelength as $\lambda_F = 2\pi/k_F \sim 42$ nm, only half of the sample thickness (~80 nm), also suggesting a relatively weak size confinement effect. This kind of system is often called wide quantum well[29, 30] which shows quasi-3D electronic structure.

In the meantime, we would like to point out that from the QHE in $Cd_3As_2$ nanoplates, the minima in $R_{xx}$ do not reach zero. Even in the $\nu = 10$ valley of sample #6 with a large activation energy $\Delta E = 10.3$ K and a well-defined plateau, the resistance minimum deviates from the activation behavior, which adopts linear relation with $1/T$ when plotted in log-scale, and saturates around 4.4 Ω (Fig. 5b), a value almost one fifth of the zero field resistance (23.3 Ω). If we remove the resistance residue from the overall value, the whole trend will fall into a relatively linear relation as shown in the inset of Fig. 5. This unique feature suggests that there may be extra scattering or other channels to prevent the system from developing perfectly localized states. Meanwhile, although we only subtract a constant value here, this resistance residue could adopt certain temperature dependence, which may be the origin of the slight deviation at high temperatures.

**Discussion**

Having extracted important information of the electronic states from the transport data, we now discuss the origin of the observed QHE. Inspired by the general idea in related systems such as topological insulators[31, 32], one plain explanation is the surface states of $Cd_3As_2$. $Cd_3As_2$ is a typical Dirac semimetal that holds double Fermi arcs on each surface[19]. Since there are plenty of density of states on surfaces due to the non-zero Fermi arc length when approaching the Dirac points[33], it is possible that the MR and Hall effect are dominated by the surface states with electrons mainly accumulating on two surfaces. Previous study[18] and our results show that even in high-carrier-density samples with thickness around one or two hundred nanometers, the surface states already have a major contribution to the SdH oscillations under perpendicular magnetic field. Hence the QHE of surface states could be observed as long as the bulk conduction is weak. This is similar to the QHE forming in the gate-confined 2D electron gases in the interface of conventional semiconductors, in which the bulk is still conducting[34]. But in our case, the bulk state is metallic and should contribute to a weak but non-negligible conductance which leads to the observed non-vanishing MR.

It is important to note that the energy of Weyl orbit involves both contributions from the surface Fermi arcs and the bulk state.[17, 35] When the Fermi level is not located at the Weyl nodes, electrons travelling through the Fermi arc will fall into the projection of bulk Fermi surface before sliding all the way to the Weyl node.[36] In order to induce the quantized transport, the involved bulk state needs to form discrete energy levels as well. In our case, the wide quantum well structure results in a series of sub-bands, among which only a few bands are occupied owing to the low carrier density. The double Fermi arcs[17] in $Cd_3As_2$ gives a degeneracy of 2. Therefore, the sub-band number should be 2 considering that there is a four-time difference between the LL index and the filling factor. This value is indeed reasonably close to the upper limit of the sub-band number (3.6 for sample #7) estimated by $2L_z/\lambda_F$, where $L_z$ is the sample thickness. Owing to the weak quantum confinement, these two sub-bands are actually very close to each other. Furthermore, the bulk propagating process only gives an additional phase term in the Lifshitz-Onsager quantization relation.[35] Hence, the Weyl orbits from nearby sub-bands can be regarded as degenerated as long as the phase difference is much smaller than the inverse of the magnetic field.

By combining several samples (Supplementary Table 1), we plot the oscillation frequency $F_\perp$ with out-of-plane magnetic field ($\theta=0°$) as a function of the square of Fermi wave vector $k_F^2$ in Fig. 5c. Here $k_F$ is the Fermi wave vector of each individual valley calculated from the oscillation frequency $F_\parallel$ with in-plane magnetic field ($\theta=90°$). Compared with that of sample #2 and #5, the value of $F_\parallel$ experiences a jump in sample #4, #6 and #7 while $F_\perp$ only slightly increases while $F_\perp$ only shows slightly increase, owing to the merging of two valleys after Lifshitz transition (Supplementary Fig. 7 and 8, Supplementary Note 3). As the Fermi level gradually passes through the Lifshitz energy, the cross-section area of bulk Fermi surface along [$1\bar{1}0$] direction ($\theta$=90°) will experience an abrupt increase and is doubled at the transition point since the two Dirac points is along the [001] direction.[19, 23] From the oscillation frequency of the QHE, the carrier density that contributes to the QHE can be calculated through $F_\perp = \frac{\pi\hbar}{2e} n_{2D}$. Here $\hbar$ is the reduced Planck constant. Taking sample #7 as an example, the carrier density from QHE is $n_{2D} = 2.2 \times 10^{12}$ cm$^{-2}$. On the other hand, the overall carrier density calculated from the Hall effect is $2.9 \times 10^{12}$ cm$^{-2}$, suggesting that the overall conduction is dominated by the carriers in the quantum Hall state. We can also calculate its bulk carrier density $n_{3D} = 2.3 \times 10^{17}$ cm$^{-3}$ from $F_\parallel/2 = \frac{\hbar}{2e}(3\pi^2 n_{3D}/2)^{2/3}$ based on a 3D Fermi surface with Fermi level slightly above the Lifshitz energy (*i.e.*, $F_\parallel$ represents the overall area of two circles with a small overlapping as shown in Supplementary Figure 7a). If following the surface-state scenario discussed above, the residue carrier (apart from those in the quantum Hall state) in the Hall effect should be these bulk carriers. Then we can infer an effective thickness of ~30 nm for the bulk state regime from the ratio between the sheet carrier density and 3D carrier density given by the SdH oscillation frequency, therefore obtaining a penetration depth of ~25 nm for each surface state (The overall thickness of sample #7 is ~80 nm). Such a penetration depth value agrees with our calculations on the finite size effect of Cd$_3$As$_2$ (Supplementary Fig. 9 and Supplementary Note 4) and other numerical simulations based the low energy model of Dirac semimetals[37]. Ignoring the curvature of Fermi arcs, the effective Fermi surface size of Weyl orbit can be simply estimated through the formula $2k_F k_0$, where $k_0$ is the projected length of the Weyl node separation on the corresponding surface.[17, 18] Consistent with the previous work on Weyl orbit[18], the overall trend of experimental data is Fig. 5c fits well with this calculation (blue curve).

Strictly speaking, in such a wide quantum well system, the bulk carriers may also give rise to QHE when tightly confined by high magnetic field. Out-of-plane magnetic field may force electrons in nanostructures to conduct 2D transport because of the small cyclotron radius (~15 nm at 10 T) due to the small Fermi wave vector and large mean free path in our $Cd_3As_2$ nanoplates (samples #2~7). However, in this scenario, there should be another component of 2D oscillations from surface state other than the 2D bulk states since the surface state contribution is enhanced when lowering the Fermi level (The Fermi surface of Weyl orbit remains finite at zero energy[36] while the bulk Fermi surface is reduced to the Weyl nodes). According to our calculations (Supplementary Fig. 9), the surface state should persist down to a few nanometers although it may no longer stay as Fermi arcs, regardless of the sub-band forming. The coexistence of 2D and 3D oscillations (Fig. 2b) at large angles also suggests the presence of multiple modes rather than simply the bulk state being confined. We further plot the relationship between $F_\perp$ with $k_F^2$ in Fig. 5c (red curve) given by this confined-bulk scenario. The original quasi-3D bulk carriers develop a 2D Fermi surface under high magnetic field with the total amount of carriers fixed. It is clear that the surface-state scenario (the blue curve) provides a better description of the data.

Meanwhile, the double Fermi arcs in Dirac semimetals have been predicted to be unstable in the presence of perturbations.[17, 36] It is possible that the surface states of $Cd_3As_2$ nanoplates are deformed into Fermi pockets rather than keeping as Fermi arcs. Although the observed large Zeeman effect can split the overlapping Weyl nodes and helps to preserve the Fermi arcs (Supplementary Note 5), there is no direct evidence for the bulk propagating process in the current results. The oscillation frequency (or Fermi surface area) is mainly determined by the Fermi arcs, while the propagation through the bulk states gives an additional phase shift.[35] Further experiments are being pursued to investigate the effect of bulk propagating part in the quantum Hall state. It is also of vital importance to investigate whether there is any fundamental difference between the exotic quantum Hall effect in Weyl orbit with that from the conventional two-dimensional electron systems.

In conclusion, we report the evolution of Weyl orbits in $Cd_3As_2$ nanoplates. When lowering the carrier density, the surface state gradually dominates the transport under perpendicular magnetic field, which further develops into an unconventional quantum Hall state. Our work reveals a potentially new quantum Hall state in 3D topological semimetals.

**Methods**

**Material growth**

The $Cd_3As_2$ nanoplates and bulk crystals were grown using $Cd_3As_2$ powders as the precursor in a horizontal tube furnace with argon as a carrier gas. The growth condition is similar to previous studies[13, 38] with deposition temperature around 150~250 °C and argon pressure around 20~200 torr. Here the carrier density of the as-grown nanoplates, mostly due to arsenic vacancies, is found to be affected by the growth rate. The modulation of parameters like higher deposition temperature and lower pressure tends to achieve low-carrier-density samples. The largest crystal plane of as-grown $Cd_3As_2$ nanoplates is (112).

**Device fabrication and transport measurements**

$Cd_3As_2$ nanoplates with thickness around 80~150 nm are chosen for this study to ensure the phase coherence of the cyclotron carriers when propagating in vertical direction. The Hall bar devices were fabricated by electron beam lithography technique and wet-etched by standard buffered HF solution for 1~3 s at the electrode regime. The electrodes were formed by Cr/Au (5nm/150nm) bilayers. The low field magnetotransport was carried out in a Physical Property Measurement System (Quantum design) with low frequency AC (<100 Hz) or DC current. The high field magnetotransport of $Cd_3As_2$ nanoplates was measured under steady high magnetic field to avoid the influence of eddy current.

**Acknowledgements**

This work was supported by the National Key Research and Development Program of China (2017YFA0303302) and National Natural Science Foundation of China (61322407, 11474058, 61674040). We thank Andrew Potter, Haizhou Lu, Daniel Bulmash, Xiao-Liang Qi, and Jing Wang for helpful discussions, Zhigang Chen for assistance in the nanostructure growth, and Mehdi Kargarian and Yuan-Ming Lu for helpful correspondence. Part of the sample fabrication was performed at Fudan Nano-fabrication Laboratory. Part of transport measurement was performed at the High Magnetic Field

Laboratory, CAS. A portion of this work was performed at the National High Magnetic Field Laboratory, which is supported by National Science Foundation Cooperative Agreement No. DMR-1157490 and the State of Florida. Stefano Sanvito thanks Science Foundation of Ireland (grant No. 14/IA/2624) for additional financial support. Computational resources were provided by the Trinity Centre for High Performance Computing.

**Figure caption**

**Figure 1| Formation of Weyl orbits and new 2D Fermi surface in $Cd_3As_2$ nanoplates.** (**a**) Illustration of Weyl orbits in $Cd_3As_2$ with the magnetic field perpendicular to the sample surface. Each set of Weyl orbit consists of two half cyclotron processes in the momentum space of the top and bottom surfaces and a real-space propagation between surfaces. The two ends of Fermi arcs connect bulk Weyl node pairs with the opposite chirality (red and blue). (**b**) Illustration of Fermi surface when the Fermi energy is not at Dirac points (left) and the geometry of magnetotransport measurement setup (right). The two Fermi surfaces are connected by the two Fermi arcs. $\theta$ is defined as the angle between magnetic field and the normal direction of the sample surface. (**c-d**) Comparisons of MR in $Cd_3As_2$ nanoplate (sample #1, **c**) and $Cd_3As_2$ bulk crystals (**d**), respectively. The insets in **c** and **d** are the extracted oscillations in sample #1 and $Cd_3As_2$ bulk crystals, respectively. (**e-f**) Comparisons of the FFT spectra of quantum oscillations in sample #1 (**e**) and $Cd_3As_2$ bulk crystals (**f**), respectively. The insets in **e** and **f** are the angle dependence of oscillation frequency in sample #1 and bulk $Cd_3As_2$ crystals, respectively.

**Figure 2| SdH oscillations in $Cd_3As_2$ nanoplate with low Fermi level.** (**a-b**) The extracted SdH oscillations of sample #2 at different angles (**a**) and LL spectrum (**b**). (**c**) Landau fan diagram in the low field regime. (**d**) The angle dependence of normalized oscillation frequency ratio. $F_s$ denotes the 2D oscillation frequency and $F_{s0}$ represents the 2D oscillation frequency at $\theta=0°$. The inset is a sketch of the magnetotransport geometry. (**e**) Comparison of $\Delta R$ and $-\frac{d^2R}{dB^2}$ curve. Two extra hidden peaks are revealed by $-\frac{d^2R}{dB^2}$ while being less obvious in the $\Delta R$ curve.

**Figure 3| Quantum Hall effect in $Cd_3As_2$ nanoplate.** (**a**) Magnetic field dependence of $R_{xx}$ (blue)

and $R_{xy}$ (red) in sample #7. (**b**) $R_{xx}$ (blue) and $R_{xy}$ (red) as a function of $1/B$. The red dash lines give the original LL positions assuming no splitting. The inset is the corresponding Landau fan diagram, which provides the LL index $N$ for each oscillation peak.

**Figure 4| Temperature dependence of $R_{xx}$ in Cd$_3$As$_2$ nanoplate.** (**a-b**) Longitudinal magnetoresistance $R_{xx}$ at different temperatures in sample #6 (**a**) and #7 (**b**). (**c**) The Arrhenius plots of the resistance minima for each Landau level to extract the activation energy through linear fitting.

**Figure 5| Energy dependence of SdH oscillation frequency.** (**a**) The extracted SdH oscillations with in-plane magnetic field in sample #6, indicating the existence of 3D bulk states. (**b**) The resistance minima of $\nu = 10$ in sample #6 at different temperatures. It gradually deviates from the thermal activation behavior and becomes saturated towards zero temperature. The inset of **b** is the resistance minimum value with the residue subtracted. (**c**) The relation of the oscillation frequency $F_\perp$ with out-of-plane magnetic field ($\theta=0°$) and the square of bulk Fermi wave vector $k_F^2$. The blue and red curves describe two predicted relations between $k_F^2$ and $F_\perp$ based on the surface-state and confined-bulk-state scenarios, respectively.

**Fig. 1**

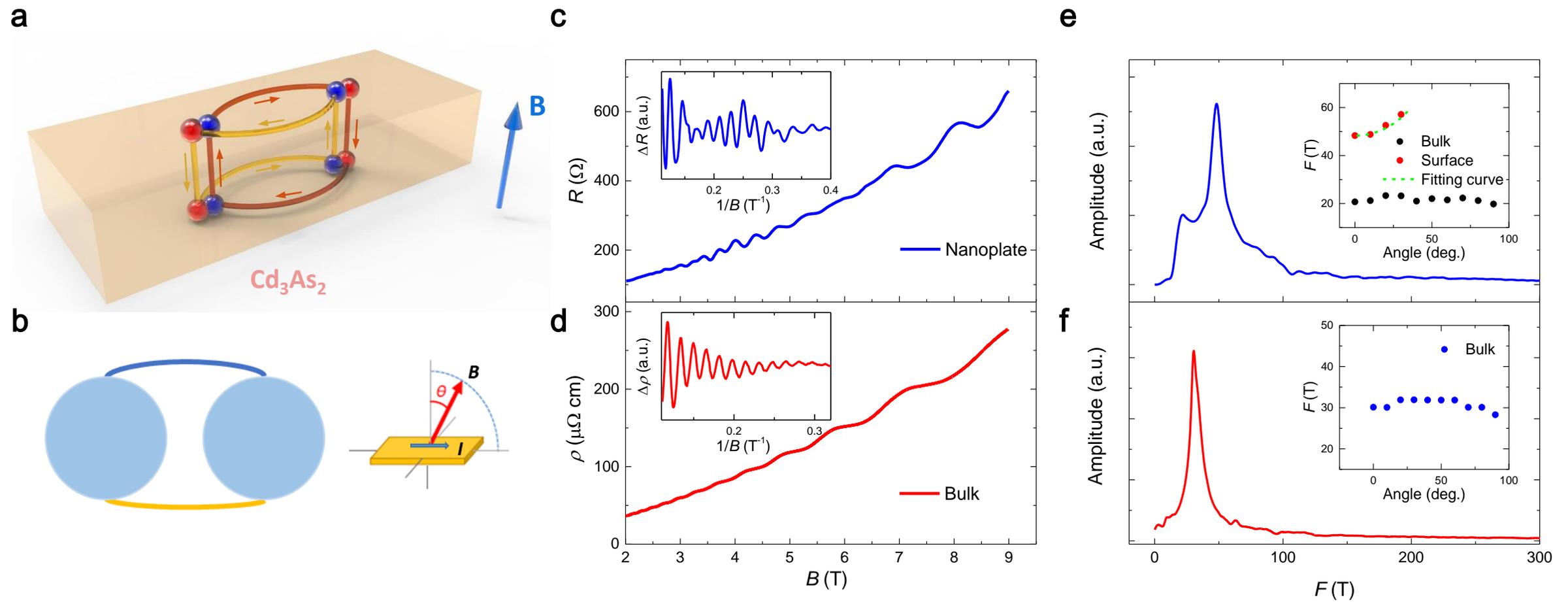

**Fig. 2**

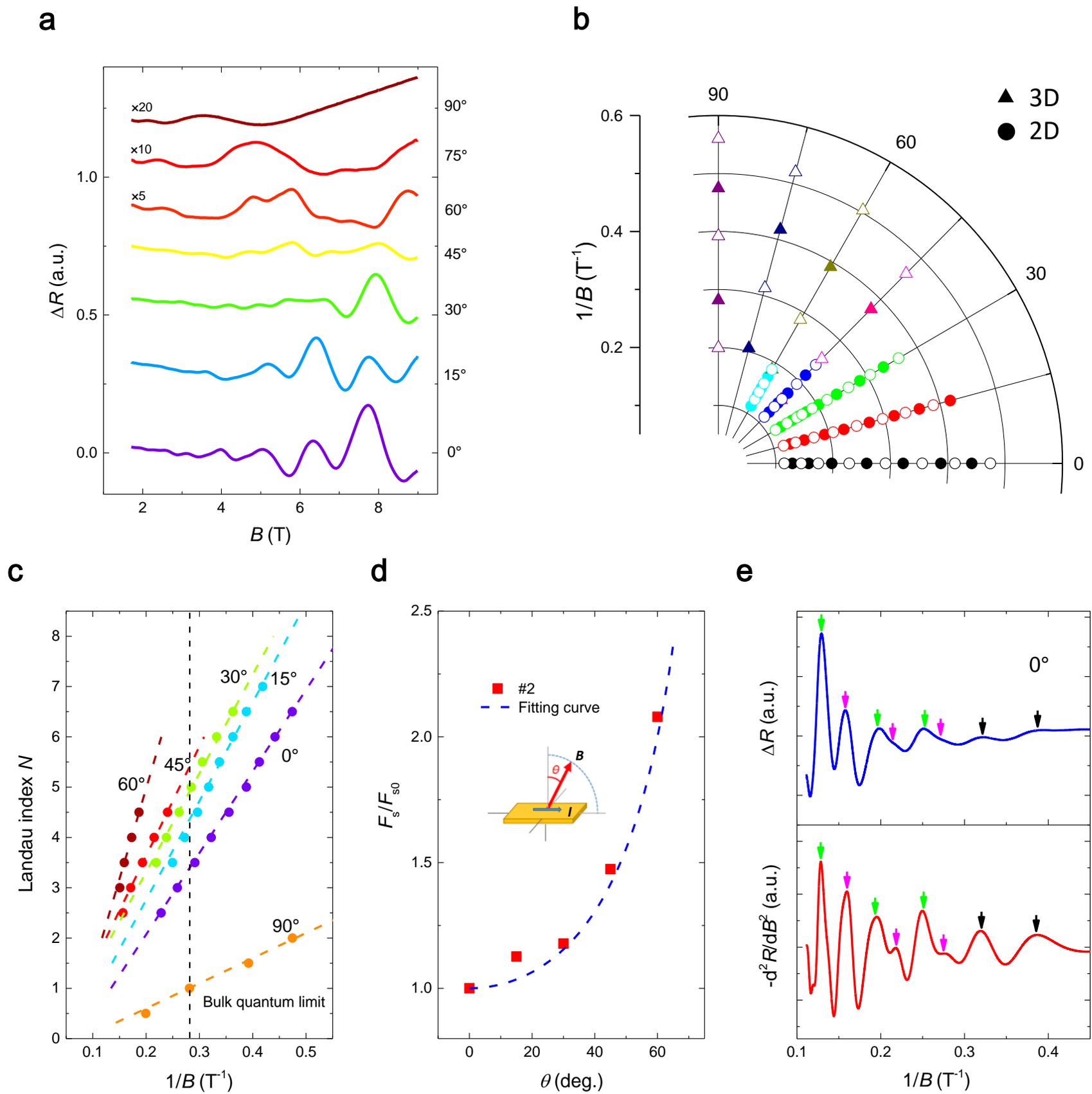

**Fig. 3**

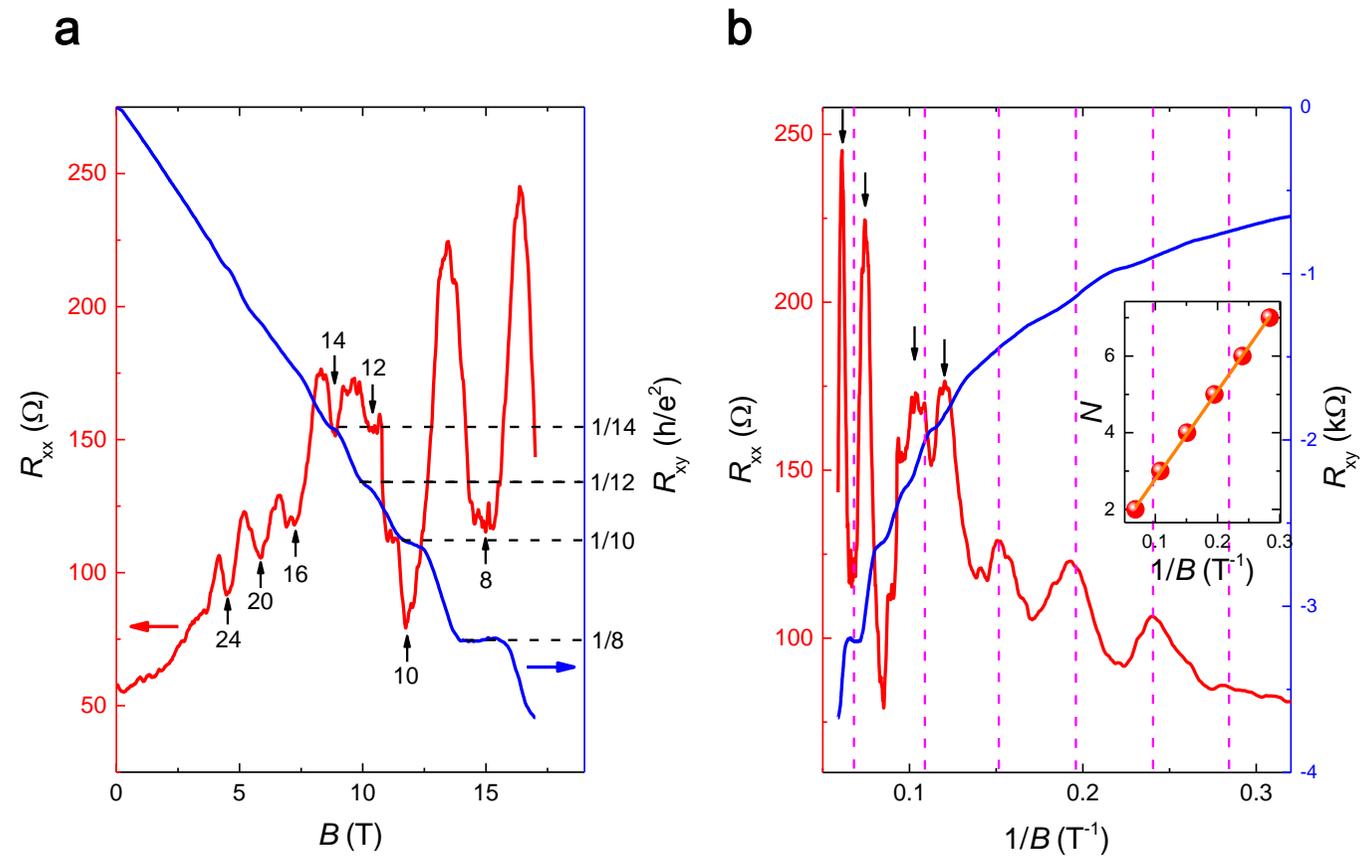

**Fig. 4**

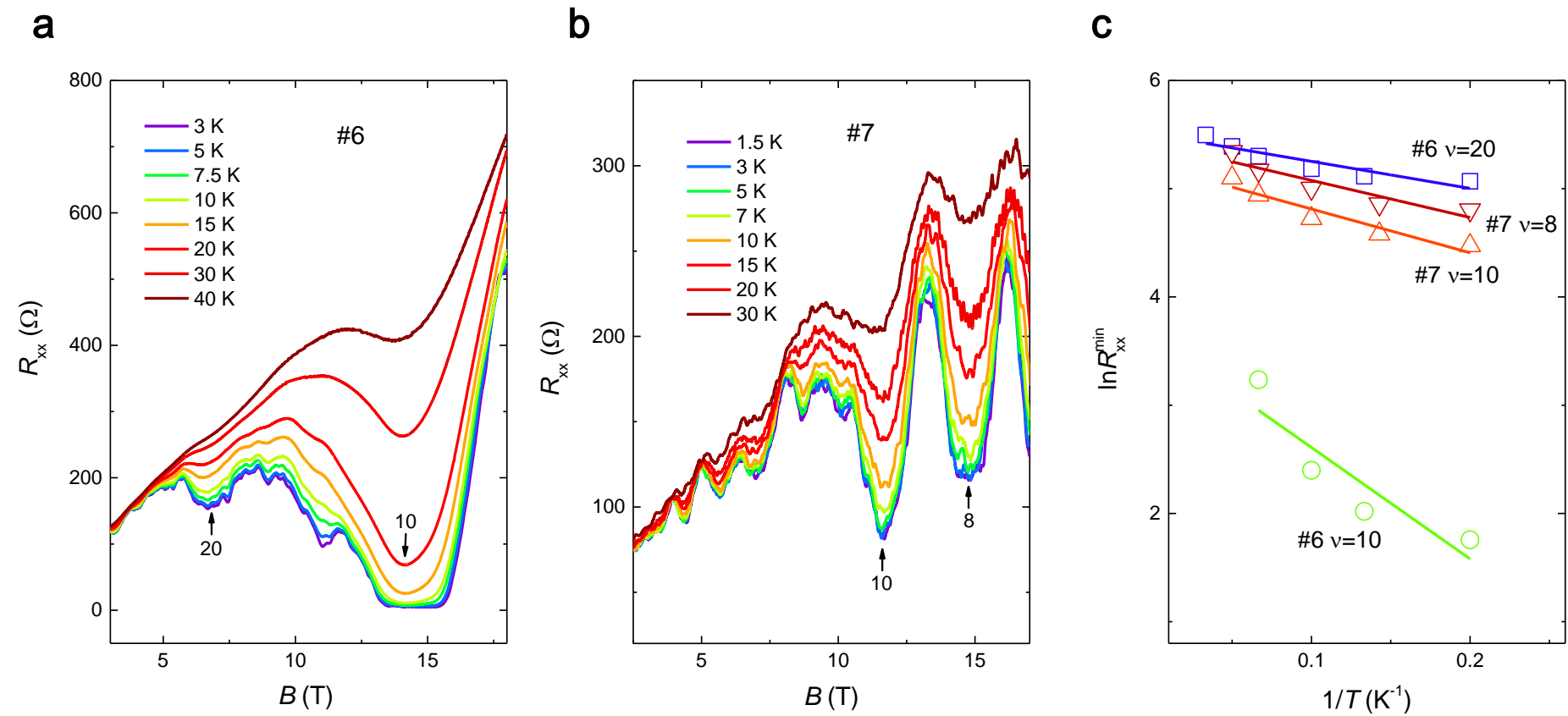

**Fig. 5**

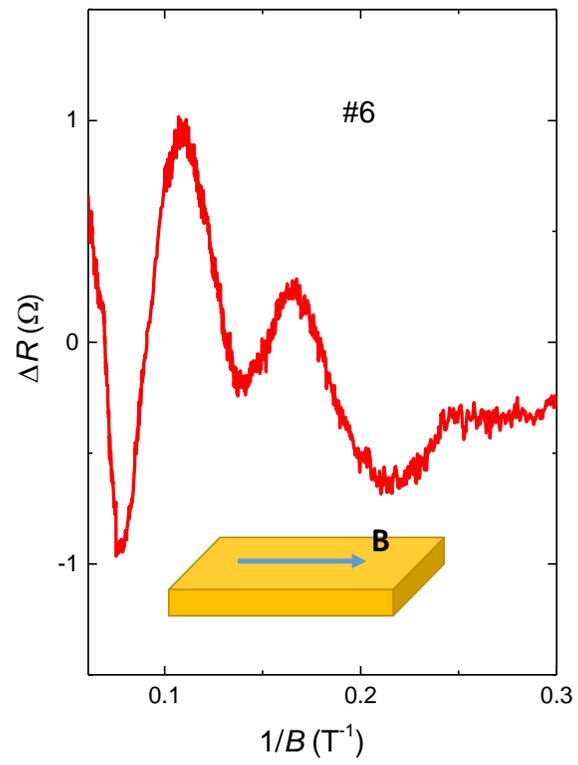 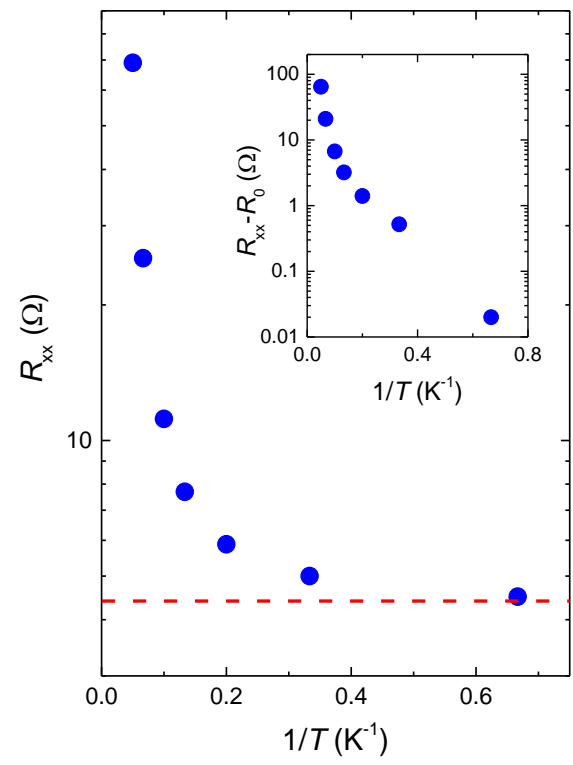 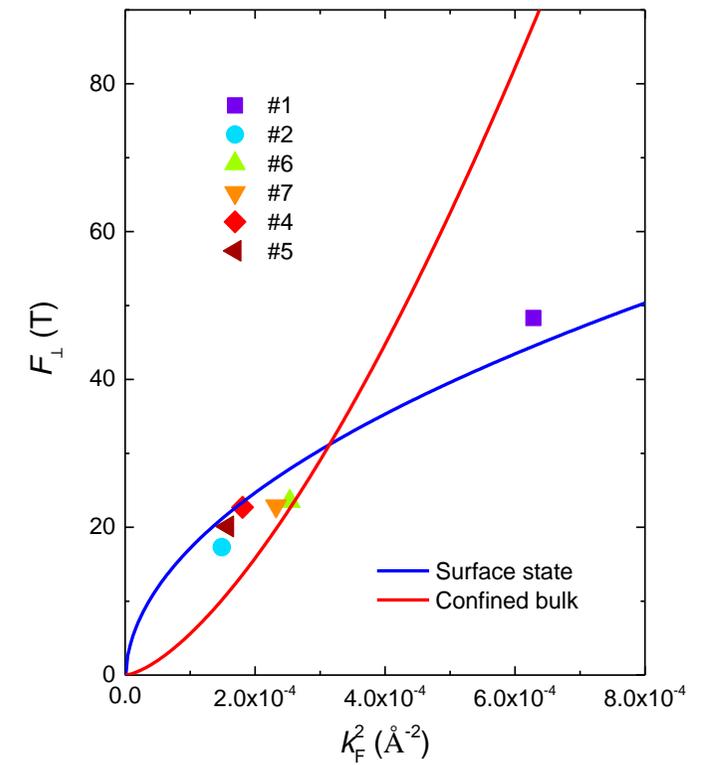